# Comparison of long-term natural aging to artificial aging in Duralumin.

*Magali* Brunet[1,*], *Benoit* Malard[2], *Nicolas* Ratel-Ramond[1], *Christophe* Deshayes[1], *Bénédicte* Warot-Fonrose[1], *Philippe* Sciau[1] and *Joël* Douin[1]

[1] CEMES, CNRS, Université de Toulouse, 29 rue Jeanne Marvig, BP 94347 31055 Toulouse Cedex 4, France
[2] CIRIMAT, Université de Toulouse, CNRS, INP Toulouse, UPS, 4 allée Emile Monso, 31030 Toulouse, France

**Abstract:** The understanding of long-term aging of aeronautical materials, in particular aluminium alloys used in the fuselage and structure of aircraft is of extreme importance for airline fleets. In this work, a plate from an old aircraft (Breguet) was retrieved and studied in terms of microstructure and mechanical properties. A comparison was made between this naturally-aged alloy and a modern alloy on which different artificial aging conditions were applied. The old alloy exhibits a precipitation of θ-$Al_2Cu$ at grain boundaries and of Ω-$Al_2Cu$ on dispersoids. This non-expected nanostructure for an alloy in T4 state was attributed to the heat that the plate experienced during the aircraft cycles. However, it is shown that this aging is reversible (after a solution treatment). Moreover, the very long time of outdoors exposure seems to have caused intergranular corrosion causing the early failure during tensile tests on some of the specimens. The artificial aging (low temperature, 100°C for up to 10,000h) applied on the modern 2017A alloy did not allow to reproduce the nanostructure of the old plate, meaning that isothermal conditions for artificial aging might not be appropriate in this case.

## 1. Introduction

Long-term behaviour of aeronautical materials is of crucial importance for industries as airline fleets are aging, with many aircraft reaching their design service goal [1]. Alteration due to fatigue and corrosion are the main concerns. The main causes come from service environment: operating temperatures, loads, ambient environmental conditions, moisture and fluid exposures, radiation, maintenance and ground handling [2]. Properties of materials degrade with time when exposed to such environment. Properties generally monitored are: mechanical static properties (yield stress, tensile strength, hardness and fracture toughness), resistance to fatigue, fatigue-corrosion and corrosion resistance [3].

The approach to evaluate the response of materials to long-term exposure in a service environment, is to impose either accelerated testing to assess mechanisms involving accumulation of damage, for example creep and high-temperature fatigue, or accelerated aging to generate end-of-life microstructure by increasing temperature or loads through furnace treatments and/or cycling. For age hardening aluminium alloys, exposure to temperature can introduce microstructural changes, whether at microscale by the formation of subgrains or at lower scale with for instance the nucleation and growth of new nanostructure phases and the change in dislocation density. These changes will lead to decrease in strength, fracture toughness or fatigue resistance.

For 2xxx alloy family, several studies were reported where long-term aging was simulated with temperature exposures [4-7]. In 1999, a study of long-term stability was done by Braun et al. [4] on 2024-T351 alloys. The authors attempted to verify whether the change of microstructure in artificially aged alloys (aging temperatures between 85°C and 150°C) had an influence on mechanical properties and corrosion behaviour. The tested sheets were aluminum clad and uncladded 1.5 mm thick sheets for fuselage. An interesting result was the fact that tensile strength, ductility and fracture toughness were not affected by the thermal exposure at 85°C for 5,800 hours, at 100°C for 720 hours and at 120°C for 3 hours. Only corrosion behaviour was deteriorated as intergranular corrosion appeared. This was related to the grain boundary precipitation occurring at these temperatures. For Al-Cu-Li alloys of the first generation in T8 state, which are known to have a unstable microstructure during long term ageing in service [8], Deschamps et al. applied low temperatures for long times: 1,000 h at 100°C or 3,000 h at 85°C [7]. They observed an evolution of the yield stress during aging at 85°C and were able to link it to the microstructure evolution, i.e. precipitation of metastable phases in the volumes between T1 precipitates.

Although accelerated aging proved to provide information on the possible evolution of materials over long-term, data from the real case are required to confirm or infirm the predictions based on artificial aging for aeronautic materials. Indeed, cycles endured by aircraft during its life-time are not isothermal. To do so, an interesting approach is to observe the microstructure and measuring the macroscopic properties of alloys retrieved on end-of-life aircraft and compare them to an equivalent alloy on which artificial aging is being conducted. There have been few attempts of such studies in the literature. Salimon at al. [9] have studied aged D16AT alloys (Russian nomenclature, equivalent of 2024-T4) retrieved from the fuselage skin of three Russian TU-154 aircraft which flew respectively 11, 15 and 20 years. They showed that for uncorroded parts, mechanical characteristics (yield stress, tensile strength and elongation to fracture) did not evolve. Low-stress, high-cycle fatigue did not cause dramatical structure changes in the bulk material. Only the near-surface layer was influenced by high-cycle fatigue, giving rise to micro-cracks and fracture of oxide film and thus corrosion. Schmidt et al. [1] studied 2024-T3 sheets taken from an Airbus 300 fuselage with 18 years of service usage (75% of its design service goal). Here too, although the old sheets exhibited a high susceptibility to intergranular corrosion, all properties tested (strength, elongation, fracture toughness, crack growth rate) were equal or even slightly better than the required properties. Our team demonstrated recently the interest in studying materials from old aircrafts (kept in museum or associations) [10]. Two plates recovered from a Breguet Sahara Deux-Ponts which first flew in 1958, were studied and compared. It appeared that the mechanical properties (yield stress, tensile strength and elongation) of these 60 years-old plates still conform to the standards of that time. Prudhomme et al. [6] compared mechanical properties of AA 2024 from a decommissioned A320 aircraft (wing panels) to an artificially aged alloy. They found similar results for an end-of-life material and an alloy artificially aged for 55h at 150°C. This aging treatment was chosen from an equivalence rule based on the Arrhenius law. The assumption is that homogeneous precipitation of second phases resulting from decomposition of a solid solution is controlled by the diffusion of copper atoms. The activation energy for copper diffusion is defined thanks to the model developed by Khan and Starink [11] for Al-Cu-Mg alloys. With this equivalence rule, 55h at 150°C should thus correspond to 100,000h exposure at 80°C, a service life of an aircraft. As a result, Prudhomme et al. showed that fatigue lives were indeed comparable as well as fracture mechanics.

---

[*] Corresponding author : magali.brunet@cemes.fr

In order to verify whether long-term artificial aging is representative of natural aging in the case of Duralumin, this article presents a comparison between an old alloy retrieved from the Breguet Sahara Deux-Ponts and a modern equivalent alloy 2017A. Different heat treatments were applied to assess the aging on the old alloy, i.e. solution treatment and artificial aging at 180°C. Long-term artificial aging at 100°C was applied to the modern alloy 2017A to simulate long-term natural aging (in-situ). Mechanical tests (hardness, tensile tests) followed by scanning electron microscope (SEM) fractography were conducted. Transmission electron microscope (TEM) was used to observe the nanostructure at different stages and to interpret the macroscopic behaviour of the artificially and naturally-aged alloys.

## 2. Experimental

### 2.1. Materials

A plate labelled L8 (1.6 mm thick) recovered from the engine nacelle of a Breguet Sahara Deux-Ponts already observed in [10] is the naturally-aged plate. A commercial 1 mm thick plate made of 2017A alloy in state T4 was bought: it is the modern alloy on which artificial aging will be conducted. On this alloy, a comprehensive study of the precipitation was previously conducted as it constitutes the first step to properly understand further aging [13].

Table 1 reports the elementary composition for each alloy measured by ICP-OES (Inductively Coupled Plasma - Optical Emission Spectrometry). The old plate is made of A-U4G (French designation used in 1950 for Duralumin). Compared with the equivalent modern alloy, it has a lower content of copper and a higher content of magnesium.

To assess aging, different aging treatments were applied, as described in Table 2. For the plate collected on the aircraft, labelled "AR" for as-received, no pre-treatment was applied. We know however from Breguet standards [14] that these pieces underwent upon fabrication a T4 treatment: solution treatment followed by water quench and room temperature aging. For this plate, the AR state corresponds thus to a T4 + 10 years of service + 50 years in outdoor conditions. The following treatments described in the table were all applied on the as-received alloys in the laboratory.

Table 1. Elemental composition of alloys obtained by ICP-OES

|  | Al | Cu | Mg | Mn | Si | Fe | Ti | Zn | Cr |
|---|---|---|---|---|---|---|---|---|---|
| **2017A-T4** | Base | 4.32 ±0.08 | 0.68 ±0.01 | 0.611 ±0.002 | 0.618 ±0.006 | 0.34 ±0.01 | 0.043 ±0.001 | 0.20 ±0.04 | 0.029 ±0.003 |
| **L8 (A-U4G-T4)** | Base | 3.88 ±0.05 | 0.83 ±0.01 | 0.522 ±0.007 | 0.60 ±0.01 | 0.53 ±0.04 | 0.063 ±0.001 | 0.045 ±0.004 | 0.023 ±0.002 |

Table 2. Condition states studied in this work

| Label | State |
|---|---|
| **AR** | As-received |
| **180°C/8h** | Thermally aged at 180°C for 8h |
| **ST** | Solution treatment: 500°C for 1h, water quenched and room temperature aged |
| **ST-180°C/8h** | Solution treatment: 500°C for 1h, water quenched and thermally aged at 180°C for 8h |
| **100°C/10,000h** | Thermally aged at 100°C for 10,000h |

### 2.2. Characterization techniques

Tensile specimens were cut in each plate: 4 in rolling direction and 4 in the transverse direction. The gauge length/reduced section and width of the tensile specimen are 20 mm and 3 mm respectively. The thickness of the 2017A plate and the L8 plate are 1 mm and 1.6 mm respectively. Tensile tests were performed at a strain rate of 0.02 s$^{-1}$, on a Zwick Z030 with a MTS extensometer. SEM pictures of fracture surfaces were taken on a JEOL 6490.

Micro hardness measurements were done with a Buehler Omnimet series 2100, following the ASTM E384 standard. Average results on 5 indents with a load of 300g during 10s were plotted. For these measurements, the alloys were cut along three perpendicular orientations: rolling direction (RD), normal direction (ND) and transverse direction (TD) and each specimen was embedded in an epoxy resin and mechanically polished on water-lubricated abrasive papers (silicon carbide) then on polishing cloths with diamond paste.

For TEM observations, the preparation was the following: to bring the sample thickness down to 25 µm, a mechanical polishing on SiC paper from 600 down to 2400 grade was performed. The specimens were then electrochemically thinned using a Tenupol-5 Struers apparatus operating at 60 V in a solution of methanol and nitric acid (3:1) at -15°C. Observations of the nanostructure were carried out on a JEOL 2010 operating at 200 kV. Bright field (BF) images were taken with an orientation slightly off the [100]$_{Al}$ or the [110]$_{Al}$ zone axis. The length of nanoprecipitates (platelets) were estimated on BF images by averaging 30 measurements.

SAXS measurements were performed on a Xeuss® equipment (Xenocs) equipped with a microsource with copper anode. Samples were thinned down to 100 µm. 3 samples-to-detector distances were considered (40 cm, 1.2 m and 2.5m).

## 3. Results

### 3.1. As-received state

#### 3.1.1. Mechanical properties

The engineering stress-strain curves for the modern plate (2017A) and the old plate (L8), both in the as-received state are shown in Figure 1 for two directions: rolling direction (RD) and transverse direction (TD). For L8, a circular plate, it was not possible to identify straightforwardly the directions [10]: a labelling D1 and D2 was thus used. They are supposedly the rolling and transverse direction respectively. Figure 1.a reveals that tensile strength curves exhibit good reproducibility for 2017A. On the old alloy however, a heterogeneous behaviour is observed. Early failure occurs quite often in direction D2, the estimated transverse direction. Table 1 summarizes the main mechanical properties. If early-failed specimens are not taken into account, the yield strength (YS), the ultimate tensile strength (UTS) and the elongation are conformed to the expectation for such alloy [15].

**Table 1.** Mechanical properties extracted from the stress-strain curves for L8 and 2017A plates in as-received state.

|  |  | E (GPa) | YS (MPa) | UTS (MPa) | True UTS (MPa) | Elongation (%) |
|---|---|---|---|---|---|---|
| L8-AR | D1 | 64±12 | 322±1 | 457±1 | 524±7 | 15±2 |
|  | D2 | 54±2 | 277±15 | 398±32 | 434±54 | 9±4 |
| 2017A-AR | RD | 57±3 | 314±1 | 439±1 | 512±8 | 17±2 |
|  | TD | 65±5 | 290±2 | 433±1 | 506±2 | 17±1 |

#### 3.1.1. Fracture surface analysis

Fracture surface analysis was carried out by SEM on post-mortem specimens. Figure 2 presents images representative of each alloy. On both alloys, equiaxed dimples were observed, typically present for ductile fracture caused by uniaxial tensile load. This type of fracture processes by microvoid coalescence. Microvoids nucleate at particles such as precipitates. After nucleation, the voids grow in the direction of the applied tensile stress. The dimples caused by coarse intermetallics are clearly visible in the microstructure of both alloys. Coarse intermetallics in 2017A-type alloys are well-identified: AlMnFeSi mostly (large, grey particles with faceted morphology) together with Al$_2$Cu (round and bright) and Mg$_2$Si (dark). They are quite numerous in both samples. Extremely small dimples were also present which could be caused by intragranular precipitation.

In the old alloy (L8), on all the early-failed specimens occurring mainly in the transverse direction, decohesive fracture along grain boundaries were observed (see framed area in Figure 2c and d.). Intergranular corrosion, unspotted before testing on the specimens, seems to be the main cause of this rupture mode. It is known that alloys of the 2xxx series are susceptible to stress-corrosion cracking and that the sensitivity depends strongly on the grain orientation. The transverse direction is thus the most susceptible to cracking, which is observed here.

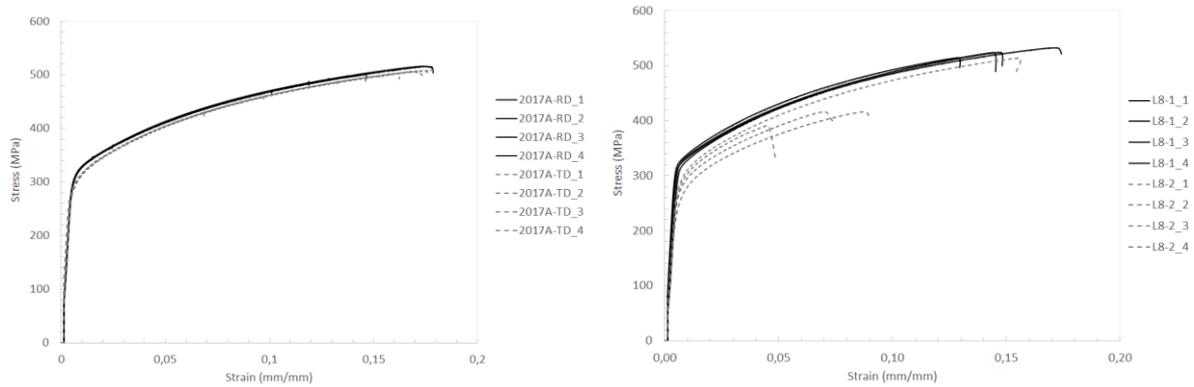

**Fig. 1.** True Stress/Strain curve in two orientations a) RD and TD for 2017A and b) D1 and D2 for L8 in as-received state.

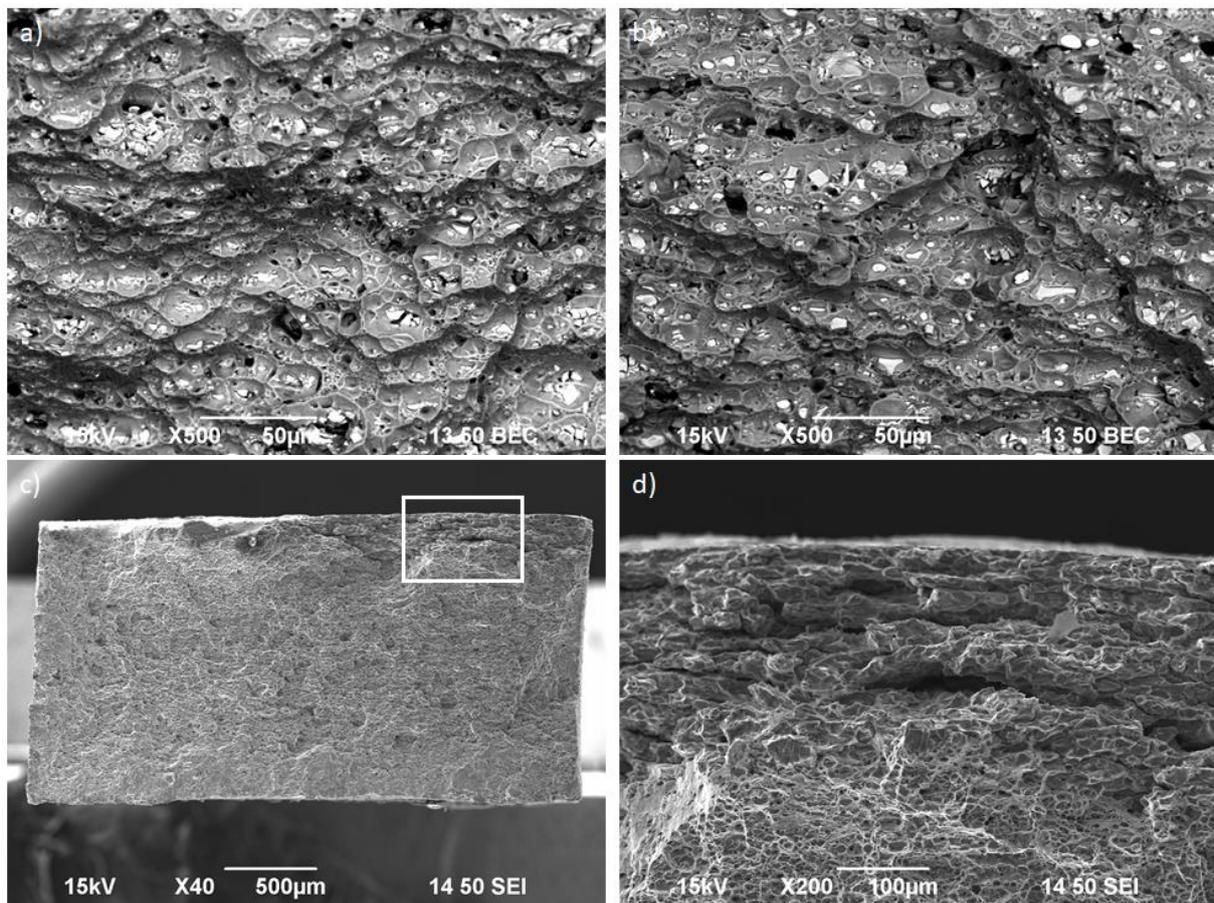

**Fig. 2.** SEM photographs of fracture surface in back-electron scattering mode for a) 2017A-AR, b) L8-AR. Images c) and d) in secondary electron mode, show fracture due to intergranular corrosion on L8-AR in the transverse direction.

### 3.2. Assessment of aging on old plate (L8)

#### 3.2.1. Microhardness

Microhardness was measured versus the aging time at 180°C for both plates, in as-received state (AR) and after a solution treatment (ST). Because the materials exhibit similar hardness in three orientation (RD, TD and ND), only hardness in RD is plotted in Figure 3. Artificial aging (180°C) was first applied on the as-received samples. For 2017A alloy, in the early stages of the treatment at 180°C, the hardness is decreasing due to the dissolution of Guinier-Preston Zones (GPZ), already present in T4 stage. Then it increases up to a maximum (peak hardening) reaching 136 HV0.3 after 8h before decreasing (over-aging stage). L8-AR exhibits a peculiar hardness behaviour upon artificial aging at 180°C. The hardness indeed does not reach a maximum after 8h at 180°C, but stays at its original value around 120 HV0.3. The attribution of such a poor behaviour in hardness to an alteration due to the history of the plate during the plane service is a possible explanation.

To verify if this behaviour is due to a long-term natural aging and to assess the reversibility of the phenomenon, the hardness of the old plate was re-measured after being solution treated at 500°C for 1 hour, quenched in water and maturated again. After this treatment, hardness of L8-ST is 123 HV0.3, close to the hardness before treatment.

Aging treatment at 180°C was then re-applied. In this case, hardness of the old alloy follows the behaviour of the alloy 2017A, reaching a maximum after 8h (see curve L8-ST in Figure 3). It can be concluded that the history of the Breguet's plate was erased. Its aging is thus reversible.

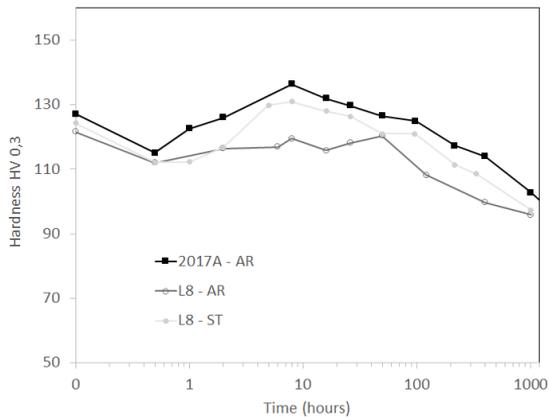

**Fig. 3.** Hardness HV0.3 measurement in the rolling direction of modern alloy 2017A and L8 (A-U4G) when aging at 180°C from an as-received state (AR) and after solution treatment (ST).

The reasons for this behaviour in the AR state are investigated with the observation of the micro and nanostructure.

*3.2.2. Microstructure*

On L8-AR, TEM observations (see Figure 4.a and 4.b) reveal a precipitation at grain boundaries and the presence of a precipitation on dispersoids. The precipitation is marked by arrows in BF-TEM image (Figure 4a) and circles in the corresponding selected area diffraction (SAD) pattern. This phase corresponds to $\Omega$-$Al_2Cu$ phase, as identified in [13] on similar Duralumin. No other precipitation is observed in this sample by TEM but it was verified by small angle X-ray scattering (SAXS) that GPZ/clusters are present in this state (see Figure 5). The results show indeed two main contributions: one in the low q-range ($0.002 < q < 0.1$ Å$^{-1}$) with a $q^{-4}$ contribution, attributed to large precipitates (dispersoids) and one at high q-range ($0.1 < q < 1$ Å$^{-1}$), a plateau attributed to GPZ or small clusters in the alloy matrix.

At grain boundaries, the precipitation nature is $Al_2Cu$. Precipitation on dispersoids and at grain boundaries could be the sign of the heat experienced by the plate during its life-time. After solution treatment (ST), precipitation on dispersoids and at grain boundaries has disappeared: see Figure 4.c and 4.d.

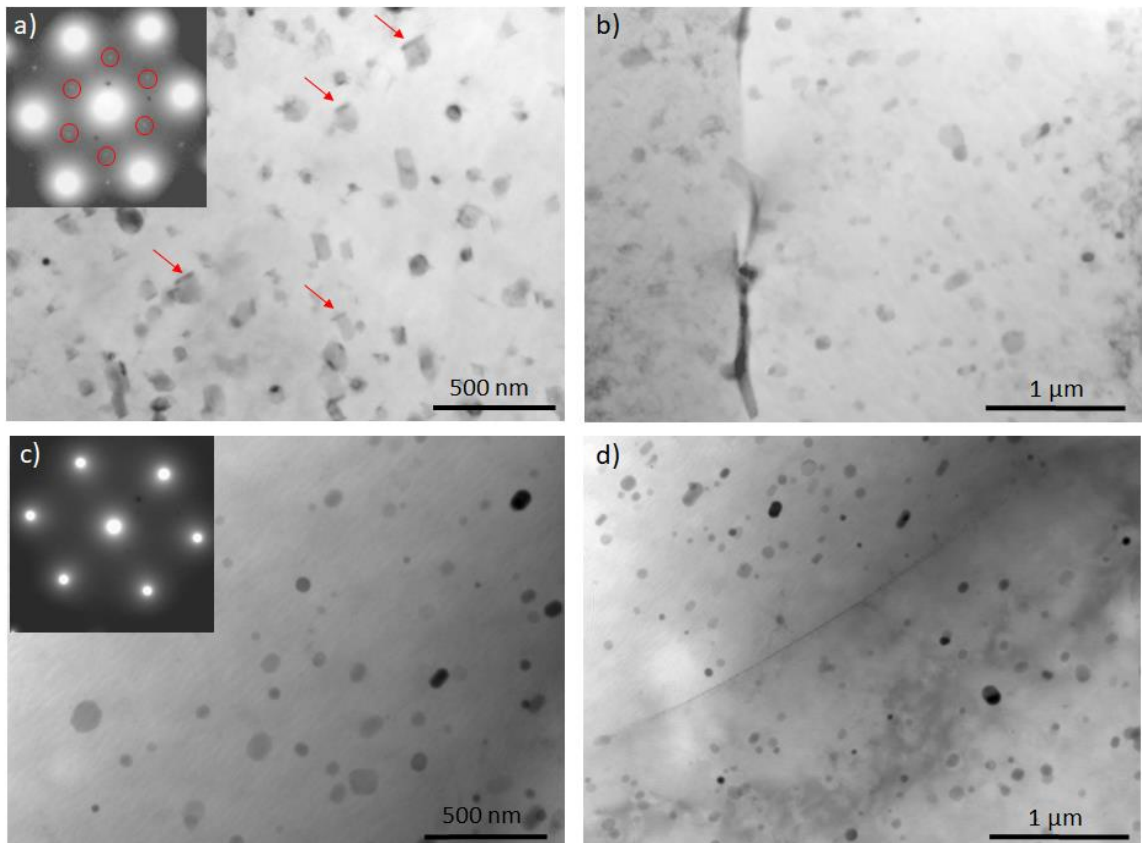

**Fig. 4.** BF-TEM images in [110]$_{Al}$ zone axis with corresponding SAD patterns of a) and b) L8-AR; c) and d) L8-ST.

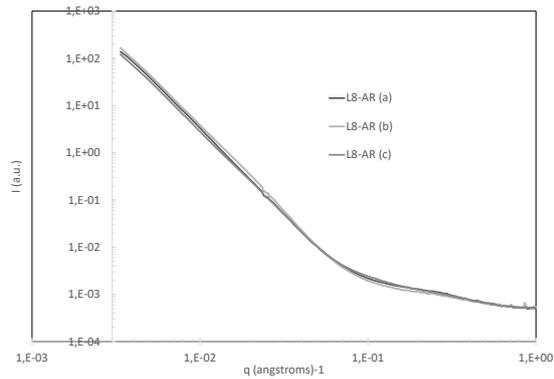

**Fig. 5.** SAXS curves for L8 (3 different zones) in the as-received state.

The nanostructure is then observed for 2017A and L8 after treatment at 180°C for 8h corresponding to peak hardening. It is shown in Figure 6. The observations of the old plate (Figure 6.a) reveals the presence of a variety of phases:
- the hardening θ''-$Al_3Cu$ phase identified in the SAD patterns thanks to the discontinuous streaks along $[200]_{Al}$ [16],
- the θ'-$Al_2Cu$ phase (platelets) identified thanks to faint spots $(110)_{θ'}$ at ½ of $(220)_{Al}$,
- the Q'(Q)-AlCuMgSi phase which are rods observed end-on in the $[100]_{Al}$ direction in the BF images,
- the Ω-$Al_2Cu$ phase with diffraction spots at 1/3 and 2/3 of the $[022]_{Al}$ orientation.

This precipitates zoology is equivalent to the one found in the modern 2017A alloy in the same aging conditions (Figure 6.b and [13]). Measurements of the hardening precipitates (θ''-$Al_3Cu$ and θ'-$Al_2Cu$ platelets) are reported in Table 3. On average, platelets are 86 ± 29 nm long for L8_180°C/8h and 60 ± 15 nm long for 2017A_180°C/8h. The difference is not significant enough to conclude. However, a much higher number of dislocations are found in the old plate indicating some work-hardening of the L8 plate during its shaping or during its lifespan.

After ST followed by aging at 180°C for 8h, the nanostructure is observed again: see Figure 6.c. The platelets are measured: in the old plate, they are 65 ± 18 nm, and dislocations have disappeared.

**Table 3.** Measurements of hardening precipitates (Platelets from θ sequence)

| Alloy | Platelets length (nm) |
|---|---|
| 2017A_180°C/8h | 60 ± 14 |
| 2017A_ST-180°C/8h | 50 ± 15 |
| L8_180°C/8h | 86 ± 29 |
| L8_ST-180°C/8h | 65 ± 18 |

These results show that hardening precipitation occurs after artificial aging (180°C-8h) in L8, in a similar manner as the 2017A alloy whether in the as-received state or after solution treatment. This means that the peculiar behaviour of the L8 in hardness during thermal aging (observed in Figure 2) is not due to a lack of hardening precipitation. The reason for the hardness stagnation on the old plate is thus suspected to be found at microscale.

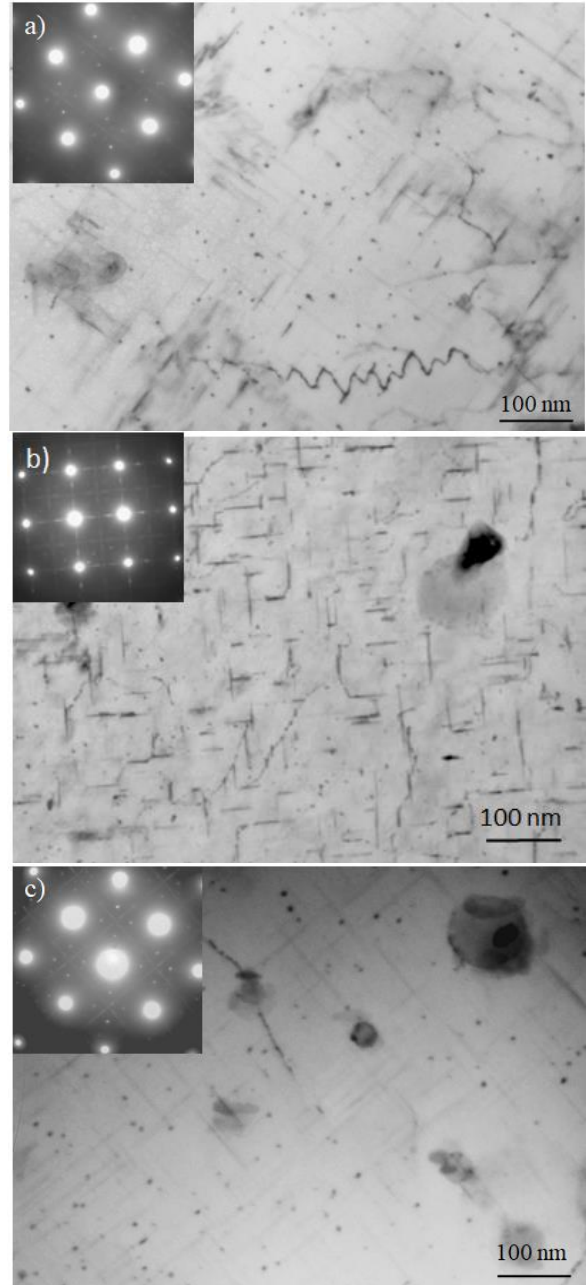

**Fig. 6.** BF TEM images in $[110]_{Al}$ zone axis a) L8_180°C/8h; b) 2017A_180°C/8h; c) L8-ST-180°C/8h

### 3.3. Long-term aging on modern alloy 2017A

As mentioned before, hardening precipitation in these alloys of the 2xxx family is considered as being controlled by the diffusion of copper atoms. The time-temperature equivalences can be used based on the copper diffusion and Arrhenius law [11]. In a recent PhD thesis work [17], several alloys among which 2024, were artificially aged at different temperatures in the range from 85°C to 250°C for different durations varying from 1 to 10,000h. Time-temperature equivalences based on microstructure evolution were found to be accurate for pre and post peak-hardening. Experimental data (mechanical properties) followed the predicted behaviour based on the evolution of the S-phase fraction.

Here, low-temperature artificial ageing (100°C) was applied on the modern 2017A alloy until 10,000h to simulate long-term aging. As hardening is based on copper diffusion, the assumption is made that the precipitation along the θ sequence should follow the same law. The time-temperature equivalence was then used with an activation energy for copper diffusion equal to 133 kJ/mol. In this case, it was calculated that the low temperature aging (100°C, 10,000h) should be equivalent to 5h at 180°C.

### 3.3.1. Microhardness measurement

For 100°C, hardness reaches 145 HV0.3 after 10,000h which is significantly higher than peak aging at 180°C (136 HV0.3). The higher hardness obtained at low temperature aging is classically observed in age-hardening alloys [18].

### 3.3.2. Microstructure

The microstructure corresponding to an aging time of 10 000h at 100°C is shown in Figure 7 and the one for 5h at 180°C in Figure 8.

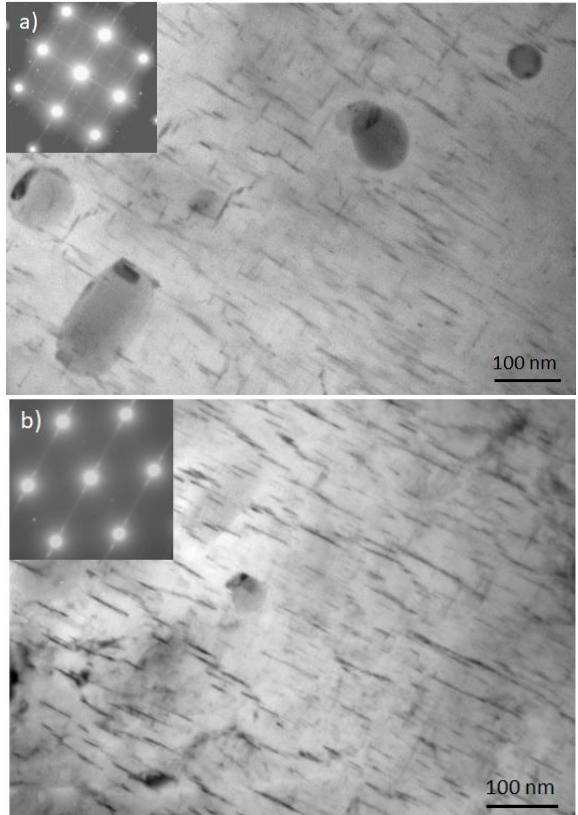

**Fig. 7.** BF-TEM images of 2017A alloy thermally aged at 100°C for 10,000h in a) $[100]_{Al}$ zone axis and b) in $[110]_{Al}$ zone axis.

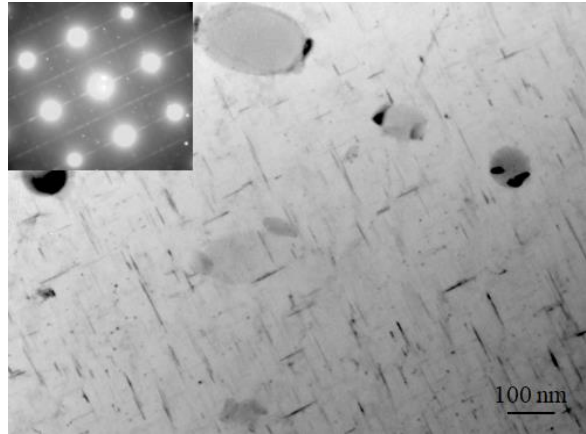

**Fig. 8.** BF-TEM images of 2017A alloy thermally aged at 180°C for 5h in $[100]_{Al}$ zone axis.

**Table 4.** Measurements of hardening precipitates (Platelets from θ sequence)

| Alloy | Platelets length (nm) |
|---|---|
| 2017A_180°C/5h | 55 ± 17 |
| 2017A_100°C/10,000h | 49 ± 21 |

In the SAD patterns in two zone axes ($[100]_{Al}$ and $[110]_{Al}$), for the 2017A alloy heat treated at 100°C for 10,000h, we observe discontinuous streaks along $[200]_{Al}$ with maximum intensities at $(100)_{Al}$ corresponding to the θ"-$Al_3Cu$ phase. The θ'-$Al_2Cu$ precipitates, if present, have a low volume density. The Q'(Q)-AlCuMgSi phase was observed in the BF images (rods seen end-on oriented along $[001]_{Al}$ direction) but the precipitates are also very few. The diffraction patterns, similar as the ones seen in [16], indicate that the alloy is just before peak hardening. Figure 8 provides the confirmation of a nanostructure very close to the one previously presented (for 8h at 180°C). The length of the hardening precipitates in the two aging conditions, is of the same order of magnitude (see Table 4). A major difference can be found for the Ω-$Al_2Cu$ phase: this phase was hardly seen in the 2017A alloy aged at 100°C for 10,000h: diffraction spots are not visible in the SAD patterns (inserts of Figure 7.a and 7.b). However, the phase can be seen on dispersoids in BF-TEM images. It is concluded that the phase is present but in very little quantities and with no significant influence.

## 4. Discussion

### 4.1. Long-term natural aging

L8 plate exhibits a peculiar mechanical behaviour and microstructure in the AR conditions, i.e. after 10 years of service and 50 years in outdoor environment. Limited elongation (or early failure) was explained by intergranular rupture mode due most probably to intergranular corrosion present in the plate. Precipitation at grain boundaries ($Al_2Cu$) was indeed observed on the AR plate by TEM. This is clearly the sign of a heat exposure during the aircraft life-time.

This precipitation induces a fragilization of the material upon exposure to environment (50 years outdoors). Another precipitation was observed in L8-AR: the $\Omega$-$Al_2Cu$ phase, which grows on dispersoids. But this precipitation has no hardening effect on the alloy in the AR state: overall L8-AR exhibits a similar hardness as 2017A. Upon artificial treatment (180°C/8h), hardness of the old plate did not reach a maximum, as opposed to the modern alloy 2017A, although hardening precipitation was observed inside the grains by TEM. We believe that microhardness is thus affected by precipitation at grain boundaries. In the AR-state, the high dislocation density observed in L8 is the sign of work hardening that compensate for the loss of hardness due to intergranular precipitation. However, this intergranular precipitation decreases the potential for further hardening in the old plate upon aging.

When a solution treatment is applied: precipitation on grain boundaries and on dispersoids ($\Omega$-$Al_2Cu$) disappear. The solution treatment erases thus the long-term aging of the plate. Consequently, the L8 behaves according to the modern alloy. The aging is reversible.

### 4.2. Artificial aging at low temperature

Artificial aging at low temperature (100°C) was applied on 2017A alloy to simulate natural aging. If time-temperature equivalence based on Arrhenius law is used, 10,000h at 100°C should correspond to 5h at 180°C. However, differences in hardness and in microstructure can be noticed between these two aging conditions.

The different age hardening responses at different aging temperature can be explained with thermodynamic considerations [18]. The free energy change per unit volume, i.e. the driving force for precipitate nucleation, is strongly influenced by the alloy composition and the aging temperature. It is higher for low temperature which makes the activity energy barrier to be the lowest, considering that the change of elastic strain energy and the change of interfacial energy are negligible versus aging temperature in this first phase of GPZ and $\theta$" nucleation. The maximum nucleation rate and therefore the maximum density of precipitates is achieved for lower temperatures, hence the maximum hardness for aging temperature of 100°C. From the TEM observations on the 2017A thermally aged at 100°C for 10,000h, the presence of coherent precipitates ($\theta$"-$Al_3Cu$) indicates that homogeneous nucleation is dominant. Although the observed platelets are of the same length as for the 180°C/5h conditions, the density of these precipitates was not precisely estimated but it appears higher after aging at 100°C than after aging at 180°C.

Another difference is noted between the two aging conditions: the presence of $\Omega$-$Al_2Cu$ phase was very rarely spotted after aging at low temperature (100°C). In AA2519 alloy (Al-5.6Cu-0.3Mn-0.3Mg-0.15Zr wt.%), Zuiko and Kaibyshev [19] showed that the precipitation of $\Omega$-$Al_2Cu$ phase happens right after the $\theta$"-$Al_3Cu$ phase precipitation and before the precipitation of the $\theta$'-$Al_2Cu$ phase. This is due to the free energy of the $\Omega$-$Al_2Cu$ phase, which is higher than that of the $\theta$"-$Al_3Cu$ phase. As it was shown that at 100°C/10,000h, the alloy is just before peak-hardening and the $\theta$"-$Al_3Cu$ phase is in the majority. It might require longer time at this temperature to generate the $\Omega$-$Al_2Cu$ phase.

Comparing 2017A artificially aged at 100°C with L8-AR, it has to be said that it was not possible to reproduce a similar nanostructure with a low-temperature aging. Long-term natural aging implies non-isothermal exposure (due to the aircraft take-off and landing cycles) which can indeed produce unpredicted nanostructure. The presence of the $\Omega$-$Al_2Cu$ phase on L8 might thus be explained by cycles of high-low temperatures experienced by the plate near the engine. Another hypothesis could also be found in the alloy composition itself: the slightly higher content of magnesium in L8 (0.83 wt.%) compared with that of the 2017A (0.68 wt.%) could help trigger this phase in L8 (always upon a heat treatment). It was shown indeed by Zuiko and Kaibyshev [19] that magnesium enhances the precipitation of transient phases.

## 5. Conclusions

From different measurements and observations at various scales on two aluminium alloys, an old one (L8) and a modern one (2017A), the following conclusions can be drawn:

- The static mechanical behaviour remains globally unchanged: upon deformation, L8 behaves similarly as 2017A alloy except for some specimens affected by intergranular precipitation and probably intergranular corrosion which caused early fracture.
- The hardness of L8 in the AR state, although close to 2017A's one, is not following the same hardening curve upon aging at 180°C. However, TEM studies reveal that hardening precipitation in L8 occurs in a similar manner as in 2017A.
  A solution treatment (500°C/1h followed by a quench) allowed to erase the particular microstructure on L8-AR, that is precipitation at grain boundaries and on dispersoids. After this, the hardness follows a typical curve with a peak-hardening after 8h.
- Time-temperature equivalence is not entirely confirmed between 180°C and 100°C: hardening precipitation (platelets) is of the same order of magnitude in length but hardness is different. This is explained by a higher volume fraction of precipitates at 100°C. Furthermore, the $\Omega$-$Al_2Cu$ phase is hardly present at 100°C: this aging temperature is thus below the $\Omega$ solvus.

- Long-term artificial aging on modern alloy was not able to reproduce the aging of the naturally-aged alloy (in-situ). This is due to the non-isothermal nature of the aging in the real case.

**Acknowledgement**

The authors wish to thank Pierre Roblin from FR FERMAT, Université de Toulouse, CNRS, Toulouse, France, for SAXS measurements and the volunteers of *Les Ailes Anciennes Toulouse* for providing us with the plate collected on the Breguet Sahara 765 aircraft. This study has been partially supported through the grant NanoX n° ANR-17-EURE-0009 in the framework of the "Programme des Investissements d'Avenir".